\documentstyle[preprint,pre,aps]{revtex}
\begin{document}
\preprint{TNT 94-2}
\draft

\title{ Effective action method for the Langevin equation}

\author{Andrea Crisanti}
\address{Dipartimento di Fisica, Universit\`a ``La Sapienza'',
         P.le A. Moro 2, I-00185 Roma, Italy}

\author{Umberto Marini Bettolo Marconi}
\address{Dipartimento di Matematica e Fisica, Universit\`a di Camerino,
         Via Madonna delle Carceri,I-62032 , Camerino, Italy}

\date{June 2, 1994}

\maketitle

\begin{abstract}
In this paper we present a formulation of the nonlinear stochastic
differential equation which allows for systematic approximations. The
method is not restricted to the asymptotic, i.e., stationary, regime but
can be applied to derive effective equations describing the relaxation of
the system from arbitrary initial conditions.
The basic idea is to reduce the nonlinear Langevin equation to an equivalent
equilibrium problem, which can then be studied with the methods of conventional
equilibrium statistical field theory. A particular well suited
perturbative scheme is that developed in quantum field theory by Cornwall,
Jackiw and Tomboulis.
We apply this method to the study of $N$ component Ginzburg-Landau
equation in zero spatial dimension. In the limit of $N\to\infty$ we can solve
the effective equations and obtain closed forms for the time evolutions of the
average field and of the two-time connected correlation functions.
\end{abstract}


\section{Introduction}
Non linear stochastic differential equations (SDE) in one or more variables
occur frequently in the description of a wide variety of physical phenomena
such as those involving relaxation towards a steady state
\cite{Schuss,Mason}.
A celebrated
example of SDE is the Langevin equation, where one relates the rate of
change of some physical observables to a drift term, i.e. to a deterministic
driving force, plus a stochastic noise.
Whereas stochastic linear differential equations are in principle amenable
to analytic solutions, those of nonlinear type are much more difficult
to treat and in most cases one has to resort either to computer simulations
or to approximate schemes.
Several approximations exist in the literature for the analysis of
nonlinear SDE. A wide used method is the so called statistical
linearization \cite{StatLin} which consists of the replacement of the
nonlinear SDE by equivalent linear ones whose coefficients are determined
by some error minimization algorithm. This method in its original
form is appropriate in the asymptotic, i.e., stationary, regime.
In many physical systems, however, it is the transient regime in which
one is interested. To deal with this problem the idea of statistical
linearization has been extended to include a possible time dependence in
the parameters \cite{West,Indiani}. The drawback of these methods is that
they are not derived in a systematic way so it is not simple to improve
them.

An alternative procedure is the so called dynamical Hartree approximation.
This scheme amounts to neglect correlations of higher order than the second
and is exact whenever the probability distribution associated with
the problem is Gaussian. This is not the case of strongly interacting
systems, so corrections to the Hartree approximation should be taken into
account. Also in this case it is not simple to improve the approximation.

In this paper we present a formulation of the nonlinear SDE which allows
for systematic approximations. This is achieved by reducing the nonlinear
Langevin equation to an equivalent equilibrium problem, which can be
analyzed with the methods of conventional equilibrium statistical field
theory. In particular we have applied a method originally developed in
quantum field theory by Cornwall, Jackiw and Toumboulis
\cite{Cornwall,Haymaker}, alternative to conventional perturbation theory,
because a normal coupling constant expansion can only be used for the study
of small corrections to the deterministic result.
In this respect the present approach is alternative to the field theoretical
treatment based on the introduction of fermionic auxiliary fields.

To illustrate the method we shall study an $N$ component Ginzburg - Landau
equation
in zero spatial dimension with the purpose of deriving systematically the
time dependent Hartree equations and the first corrections.
The same model has been discussed previously by Scalapino and coworkers
\cite{Scalapino}, who introduced an approximation scheme for the Langevin
equation based on the expansion in the small parameter $1/N$. We shall
obtain a solution which represents a systematic treatment, and as a bonus
we derive a closed equation for the two-time connected correlation
function. This will be discussed in Sects. \ref{sec:model},
\ref{sec:solution} and \ref{sec:bHartree}, where the corrections
to the $N\to\infty$ solution are considered.

The formalism is established in Sec. \ref{sec:formalism} where we
construct the generating
functional $\Gamma$ for the average value of the observables and its
correlations.
In Sect. \ref{sec:variational} we provide an alternative  derivation of
the Hartree equations, based on a variational principle similar to the
Feynman method in equilibrium statistical mechanics, and discuss the
relevant case $N=1$.

\section{The formalism}
\label{sec:formalism}
In this Section we derive the effective action formalism for the Langevin
equation. The path integral method constitutes a
convenient representation of the Langevin
equation for a field $\phi(t)$. Within this approach, the original stochastic
differential equation, where the $\phi$ depends on another field $\xi$,
called the noise, is reformulated by constructing an effective action
for the field $\phi$ only obtained by eliminating the noise.
The advantage of this
transformation is that one can employ the well known methods of
equilibrium statistical field theory.
To keep the notation as simple as possible
the derivation will be carried out for a single component real
field. The extension to $N$-vector fields will be
discussed later.

The time evolution of the field $\phi(t)$ is governed by the Langevin
equation:
\begin{equation}
\label{eq:Lang}
   \frac{\partial\phi}{\partial t} =
               - \frac{\partial}{\partial\phi}\, S[\phi]
               + \xi
\end{equation}
where $S[\phi]$ is an ``energy'' function, and $\xi$ a Gaussian random
variable with
\begin{equation}
\label{eq:noise}
  \langle \xi(t) \rangle = 0, \qquad
  \langle \xi(t)\,\xi(t') \rangle = \Gamma\,\delta(t-t').
\end{equation}
We shall derive a generating functional from which the correlations
can be obtained. Proceeding in the standard way, see e.g. Ref.
\cite{Zinn}, we introduce an external source $J$ and define the generating
functional
\begin{eqnarray}
   Z[J] = {\cal N}\,\int {\cal D}''\phi\, {\cal D}\xi\,&& {\cal P}[\phi(0)]\,
          \delta(\phi - \phi_{\xi})\,
          \nonumber \\
          &&\times\exp\left[ - \int_0^{\tau}\, {\rm d} t\, J\phi\right]
          \exp\left[ - \int_0^{\tau}\, {\rm d} t\, \frac{\xi^2}{2\Gamma}
              \right]
\label{eq:gf1}
\end{eqnarray}
where $\phi_{\xi}$ is the solution of stochastic eq. (\ref{eq:Lang})
subject to some set of initial value conditions $\phi(0)$ assigned
with probability ${\cal P}[\phi(0)]$, and ${\cal N}$ is a
normalizing constant. The functional integral on $\phi$ in
eq. (\ref{eq:gf1}) includes integration over $\phi(0)$ and $\phi(\tau)$.
We denote it by the double quote: ${\cal D}''\phi$.
The $\delta$-function stands for
\begin{equation}
\label{eq:delta}
  \delta(\phi - \phi_{\xi}) = \delta\left[ \frac{\partial\phi}{\partial t}
                                + \frac{\partial S}{\partial\phi} - \xi
                                    \right]\,
      \det\left|\frac{\delta\xi}{\delta\phi}\right|
\end{equation}
where $\det|\delta \xi/\delta\phi|$ is the Jacobian of the transformation
$\xi\to\phi$. With well-known manipulations, see e.g. Ref. \cite{Zinn,Gozzi},
one has
\begin{equation}
\label{eq:det}
   \det\left|\frac{\delta\xi}{\delta\phi}\right| =
     \exp\left[ \frac{1}{2}\,\int_0^{\tau}\, {\rm d} t\,
                \frac{\partial^2 S}{\partial\phi^2}
         \right].
\end{equation}
In deriving eq. (\ref{eq:det}) we have used the forward time propagation
Green function $\theta(t-t')$ of the operator $\partial_t$. Moreover we
have used the definition
\begin{equation}
\label{eq:tet0}
 \theta(0) = \frac{1}{2}.
\end{equation}
This choice corresponds to the ``physical'' regularization of the noise
term $\xi$ as
\begin{equation}
  \langle \xi(t)\,\xi(t') \rangle = \Gamma\,\eta(t-t')
\end{equation}
where $\eta(t)$ is an even function sharply peaked at $t=0$, whose integral
from  $-\infty$ to $+\infty$ is equal to $1$. The $\delta$-correlated noise
is obtained in the limit of vanishing width. In terms of stochastic
differential equations this corresponds to the Stratonovich formalism
\cite{Risken}.

At this stage one eliminates the noise field by
inserting eq. (\ref{eq:det}) into eq. (\ref{eq:gf1}) and performing the
$\xi$ integral over the noise obtaining,
\begin{eqnarray}
   Z[J] = {\cal N}\,\int {\cal D}''\phi\, {\cal P}[\phi(0)]\,
          \exp\Bigl[
                     &-& \int_0^{\tau}\, {\rm d} t\, \frac{1}{2\Gamma}
              \left(\frac{\partial\phi}{\partial t}
                  + \frac{\partial S}{\partial\phi}\right)^2
              \nonumber \\
              &+& \frac{1}{2}\,\int_0^{\tau}\, {\rm d} t\,
                           \frac{\partial^2\phi}{\partial\phi^2}
              - \int_0^{\tau}\, {\rm d} t\, J\phi
              \Bigr].
\label{eq:gf2}
\end{eqnarray}
The argument of the exponential can be simplified by performing the
integration of the term
$
  \int_0^{\tau}\,{\rm d} t\, \dot\phi\, \partial S/\partial\phi =
	S[\phi(\tau)] - S[\phi(0)],
$
so we finally have
\begin{eqnarray}
   Z[J] = {\cal N}\,\int
          {\cal D}\phi(0)\,&& {\cal P}[\phi(0)]\, {\rm e}^{S[\phi(0)]/2\Gamma}
          {\cal D}\phi(\tau)\, {\rm e}^{S[\phi(\tau)]/2\Gamma}
          \nonumber \\
           && \times\,  {\cal D}\phi\,
          \exp\left[
                     - I(\phi)
                     - \int_0^{\tau}\, {\rm d} t\, J\phi
              \right]
\label{eq:gf3}
\end{eqnarray}
where the action $I(\phi)$ is given by
\begin{equation}
\label{eq:rgf2}
   I(\phi) =
        \frac{1}{2\Gamma}\,\int_0^{\tau} {\rm d} t\,\left[
                \left(\frac{\partial\phi}{\partial t}\right)^2
              + \left(\frac{\partial S}{\partial\phi}\right)^2
                                              \right]
       -\frac{1}{2}\,\int_0^{\tau} {\rm d} t\,
                         \frac{\partial^2 S}{\partial\phi^2}.
\end{equation}
and ${\cal D}\phi$ denotes integration over all paths starting at
$\phi(0)$ for
$t=0$ and ending at $\phi(\tau)$ for $t=\tau$. It is defined as
\begin{equation}
   {\cal D}\phi = \lim_{N\to\infty}\, \prod_{i=1}^{N-1}\, {\rm d}\phi(t_i)
\end{equation}
where $\phi(t_i)$ is the field at time $t_i= i \epsilon$, having sliced the
interval $0$ to $\tau$ in $N$ parts of size $\epsilon = \tau/N$.

In eq.(\ref{eq:gf3}) the integration over the end points just fixes the
boundary conditions at $t=0$ and $t=\tau$, so without lost of generality,
we can consider a `reduced' generating functional
\begin{equation}
\label{eq:rgf1}
   Z[J] = {\cal N}\,\int_{0,\phi_0}^{\tau,\phi_1}
           {\cal D}\phi\,
          \exp\left[
                     - I(\phi)
                     - \int_0^{\tau}\, {\rm d} t\, J\phi
              \right]
\end{equation}
where now the integral runs
over all paths which start at $t=0$ from $\phi(0) = \phi_0$ and end
at $t=\tau$ at $\phi(\tau) = \phi_1$.
In the limit $\tau\to\infty$, or equivalently $t/\tau\ll 1$, the path becomes
independent of the final value $\phi(\tau)$.

We note however that the presence of the additional constraint is necessary
to select paths which are solution of the original equation of motion
eq. (\ref{eq:Lang}).
Equation (\ref{eq:rgf2}) and (\ref{eq:rgf1}) lead in fact to second
order differential equations of motion,
whereas the original stochastic equation
is of the first order. The additional constraint at $t=\tau$
makes the problem well defined since the paths in eq. (\ref{eq:rgf1})
must satisfy the two constraints $\phi(0)=\phi_0$ and
$\phi(\tau)= \phi_1$.
Once the two boundaries conditions are imposed the
path is also solution of the first order differential equation (\ref{eq:Lang}),
as can be easily seen in the limit $\Gamma\to 0$, i.e., the deterministic
limit.

In general the calculation of path integrals like eqs.
(\ref{eq:rgf2}) and (\ref{eq:rgf1})
is not at all straightforward. Nevertheless quantities of physical interest
can be obtained. In our case we are interested into the noise-averaged value
of the field $\langle\phi(t)\rangle$ and correlations
$\langle\phi(t)\,\phi(t')\rangle$ as a functions of time.
An advantage of the present formalism is that self-consistent,
systematic variational principles for these quantities can be obtained
using a method introduced by Cornwall, Jackiw and Tomboulis
in Quantum Field Theory \cite{Cornwall}. The basic idea is to derive an
effective action which is stationary at the physical values of
$\langle\phi(t)\rangle$ and
$\langle\phi(t)\,\phi(t')\rangle$.

The method starts by generalizing eq. (\ref{eq:rgf1})
to account
for the composite operator $\phi(t)\phi(t')$. We then define the generating
functional
\begin{eqnarray}
   Z[J,K] = {\cal N}\,\int_{0,\phi_0}^{\tau,\phi_1}
           {\cal D}\phi\,
          \exp \Bigl[ - I(\phi)
                      & - & \int_0^{\tau}\, {\rm d} t\, J(t)\,\phi(t)
                      \nonumber \\
                      & - & \frac{1}{2}\int_0^{\tau}\, {\rm d} t\,
                                  \int_0^{\tau}\, {\rm d} t'\,
                                 \phi(t)\,K(t,t')\phi(t')
              \Bigr]
\label{eq:gfkj}
\end{eqnarray}
where $J$ and $K$ are a local and a bilocal source, respectively.

By taking functional derivatives with respect to the external sources
the averaged correlations of $\phi$ can be obtained. In particular,
by considering $W[J,K] = -\ln Z[J,K]$, we have for $0<s<\tau$
\begin{equation}
\label{eq:aver}
 \left\{
    \begin{array}{ll}
    \frac{\displaystyle\delta}{\displaystyle \delta J(s)}\, W[J,K] &
                       = \langle\phi(s)\rangle
                                        \equiv q(s) \\
           & \\
    \frac{\displaystyle\delta}{\displaystyle \delta K(s,s')}\, W[J,K] & =
                           \frac{1}{2}\langle\phi(s)\,\phi(s')\rangle
                     \equiv \frac{1}{2}\left[q(s)\,q(s') + G(s,s')\right] \\
    \end{array}
\right.
\end{equation}
where the averages are obtained
with the weight of eq. (\ref{eq:gfkj}). In the limit
of vanishing external sources $q$ and $G$ become the noise-averaged field
$\langle\phi(s)\rangle$ and connected two-point correlation function
$\langle\phi(s)\,\phi(s')\rangle_{\rm c}$ of the process described by the
Langevin equation (\ref{eq:Lang}).

By Legendre transforming $W[J,K]$ we can eliminate $J$ and $K$ in favor of
$q$ and $G$:
\begin{eqnarray}
   \Gamma[q,G] = W[J,K] & - & \int_0^\tau {\rm d} s\, q(s)\,J(s)
                - \frac{1}{2} \int_0^\tau {\rm d} s\,
                              \int_0^\tau {\rm d} s'\, q(s)\,K(s,s')\,q(s')
                              \nonumber \\
              & - &\frac{1}{2} \int_0^\tau {\rm d} s\,
                              \int_0^\tau {\rm d} s'\, G(s,s')\,K(s,s')
\label{eq:Ltr}
\end{eqnarray}
where $J$ and $K$ are eliminated as a function of $q$ and $G$ by the use
of eq. (\ref{eq:aver}). It can be shown that $\Gamma[q,G]$ is the
generating function of 2PI (two-particle irreducible) Green functions,
i.e. it is given by all diagrams that cannot be separated in two pieces
by cutting two lines \cite{Cornwall,Haymaker}.
The external sources can be obtained from $\Gamma[q,G]$ as
\begin{equation}
\label{eq:aver1}
 \left\{
    \begin{array}{ll}
    \frac{\displaystyle\delta}{\displaystyle\delta q(s)}\, \Gamma[q,G] &
                           = - J(s)
                                    - \int_0^\tau {\rm d} s\, K(s,s')\,q(s) \\
	& \\
    \frac{\displaystyle\delta}{\displaystyle\delta G(s,s')}\, \Gamma[q,G] & =
                          - \frac{1}{2}\, K(s,s'). \\
\end{array}
\right.
\end{equation}
The physical process corresponds to vanishing sources, $J=K=0$. From
eq. (\ref{eq:aver1}) it follows that in this limit the value of $q$ and
$G$ are determined by the stationary point of $\Gamma[q,G]$. We have thus
obtained a variational principle for the the noise-averaged field
$\langle\phi(s)\rangle$ and connected two-point correlation function
$\langle\phi(s)\,\phi(s')\rangle_{\rm c}$ of the process described by the
Langevin equation (\ref{eq:Lang}).

The next step is to evaluate $\Gamma[q,G]$. Following Ref.
\cite{Cornwall,Haymaker} $\Gamma[q,G]$ can be written as
\begin{equation}
\label{eq:2pigf}
   \Gamma[q,G] = I(q) + \frac{1}{2}\,{\rm Tr}\,\ln\, G^{-1}
                 + \frac{1}{2}\,{\rm Tr}\,{\cal D}^{-1}(q)\,G
                 + \Gamma_2(q,G) + \hbox{\rm const.}
\end{equation}
where $I(q)$ is given by eq. (\ref{eq:rgf2}) with $\phi\to q$,
\begin{equation}
\label{eq:effpro}
   {\cal D}^{-1}(q) \equiv \left.\frac{\delta^2 I(\phi)}
                                     {\delta\phi(s)\delta\phi(s')}
                          \right|_{\phi=q}
                    = D^{-1} + \left.\frac{\delta^2 I_{\rm int}(\phi)}
                                     {\delta\phi(s)\delta\phi(s')}
                          \right|_{\phi=q}
\end{equation}
with $D^{-1}$ propagator of the ``free'' theory. The functional
$\Gamma_2$ is given by the sum of all 2PI vacuum diagrams of a theory
with interactions determined by $I_{\rm int}$ and propagators $G$. The
interaction term is defined by the shifted action
\begin{equation}
\label{eq:inter}
   I(q+\phi) - I(q)
     - \phi\,\left.\frac{\delta I(\phi)}{\delta\phi}\right|_{\phi=q} =
     \frac{1}{2} \phi{\cal D}^{-1}(q)\phi + I_{\rm int}(\phi;q).
\end{equation}
This procedure corresponds to a dressed loop expansion with vertices which
depend on $\phi$, and can thus
exhibit non-perturbative effects even for a small number of dressed loops.
The crucial point is that it in no sense corresponds to a perturbation
theory in physical amplitudes. The physics is obtained by going to the
stationary point of the expansion with respect to both variables $\phi$ and
$G$. This leads to nonlinear dynamical equations for $\phi$ and $G$. If one
could sum up the whole series, the exact value of $\phi$ and $G$ will emerge
from the stationary point. If the series is truncated one gets approximate
values of $\phi$ and $G$. The bonus is that on can get equations
describing nonperturbative behaviors. The drawback is that in general
there is no systematic knowledge about errors occurred.

In the next section we apply the above formalism to a $O(N)$ problem
where the leading contribution to $\Gamma_2$ can be extracted in the
$N\to\infty$ limit.

\section{The Model}
\label{sec:model}
To illustrate the formalism introduced in the previous section we consider
an $N$-component
Ginzburg-Landau time-dependent field $\phi_i$ with
quadratic local interaction in zero spatial dimension.
When discussing fluctuations effects to any given order in a
perturbation expansion one is not usually able to justify the neglect of
yet higher orders. However for theories with large $N$ internal symmetry
group there exists another perturbative scheme, the $1/N$ expansion.
 The model is specified by the evolution equation
\begin{eqnarray}
   \frac{\partial\phi_i}{\partial t} & = &
               - \frac{\partial}{\partial\phi_i}\, S[\bbox{\phi}]
               + \xi_i
	\label{eq:LangN} \\
    S(\bbox{\phi}) & = & \frac {a}{2} \bbox{\phi}^2
  + \frac {\lambda} {4! N} (\bbox{\phi}^2)^2
        \label{eq:GinzLand}
\end{eqnarray}
where Gaussian random noise is defined by:
\begin{equation}
\label{eq:noiseN}
  \langle \xi_i(t) \rangle = 0, \qquad
  \langle \xi_i(t)\,\xi_j(t') \rangle = \Gamma\,\delta_{ij}\delta(t-t')
\end{equation}
and we assume $a<0$ and $\lambda>0$.
Generalizing eq. (\ref{eq:rgf2}) to $N$-component fields one finds the
following action
\begin{equation}
   	I(\bbox{\phi}) =
        \frac{1}{\Gamma}\,\int_0^{\tau} {\rm d} t\,\left[
               \frac{\dot{\bbox{\phi}}^2}{2} + \frac{m^2}{2}
               \bbox{\phi}^2
               + \frac{\lambda_0}{4! N} (\bbox{\phi}^2)^2
               + \frac{g_0}{6! N^2} (\bbox{\phi}^2)^3
               -\frac{a N}{2}
                                              \right]
\label{eq:hamil}
\end{equation}
where the parameters $m^2$, $\lambda_0$ and $g_0$ are related to the
original constants by:
\begin{mathletters}
\label{eq:table}
\begin{equation}
  m^2 = a^2 - \frac{\lambda\Gamma}{6} - \frac{\lambda\Gamma}{3N},
\label{eq:tablem2}
\end{equation}
\begin{equation}
  \lambda_0 = 4\,a\lambda,
\label{eq:tablel0}
\end{equation}
\begin{equation}
  g_0 = 10\,\lambda^2.
\label{eq:tableg0}
\end{equation}
\end{mathletters}
The last term in eq. (\ref{eq:hamil}) does not depend on $\phi$ and can be
absorbed into the definition of the normalizing constant ${\cal N}$ in
eq. (\ref{eq:rgf1}).

In the limit $N\to\infty$ we can calculate explicitly the leading order
term of the functional (\ref{eq:2pigf}) following the same steps of
Ref. \cite{Dominici}. The ``action'' (\ref{eq:hamil}) corresponds to
a classical $\phi^6$-theory in one spatial dimension.

{}From eq. (\ref{eq:hamil}) it follows that the leading contributions to
eq. (\ref{eq:2pigf}) for $N\to\infty$ are
\begin{equation}
\label{eq:trdg}
  \frac{1}{2}\, {\rm Tr}\,{\cal D}^{-1}\,G = \frac{N}{2}\,
    \int_{0}^{\tau}\,{\rm d}t\,\int_{0}^{\tau}\,{\rm d}t'
    \left[
       -\frac{\partial^2}{\partial t^2} + m^2
       + \frac{\lambda_0}{3!N} q(t)^2
       + \frac{g_0}{5!N^2} q^4(t)
    \right]\, G(t,t')\, \delta(t-t')
\end{equation}
and
\begin{equation}
\label{eq:dm1}
    {\cal D}^{-1}(t,t')=
      \left[
       -\frac{\partial^2}{\partial t^2} + m^2
       + \frac{\lambda_0}{3!N} q(t)^2
       + \frac{g_0}{5!N^2} q^4(t)
      \right]\, \delta(t-t').
\end{equation}
The leading order 2PI diagrams
$N\to\infty$, shown in Fig. \ref{fig:fig1}, lead to
\begin{equation}
\label{eq:g2n}
   \Gamma_2(q,G) =  \frac{N\,\lambda_0 }{4!}\,\int_{0}^{\tau}{\rm d}t\,
G^2(t,t)
                     + \frac{N\,g_0 }{6!}\,\int_{0}^{\tau}{\rm d}t\, G^3(t,t)
                     + \frac{3\,g_0}{6!}\,\int_{0}^{\tau}{\rm d}t\,q^2(t)\,
                     G^2(t,t)
\end{equation}
where $q(t) = \langle\phi_1(t)\rangle$ assuming that the symmetry is broken
along the direction `$1$'.

Stationarity of the functional $\Gamma[q,G]$ with respect to $q(t)$ and
$G(t,t')$ yields the dynamical equations for the order parameter and its
fluctuations which read, respectively
\begin{eqnarray}
  \biggl[
    -\frac{\partial^2}{\partial t^2} + m^2 &+& \frac{\lambda_0}{3!N}\,q^2(t)
    + \frac{g_0}{5!N^2}\,q^4(t)  \nonumber \\
   & + & \frac{\lambda_0}{3!}\,G(t,t)
    + \frac{g_0}{5!}\,G^2(t,t) + \frac{2\,g_0}{5!N}\, q^2(t)\,G(t,t)
  \biggr]\, q(t) = 0
\label{eq:op}
\end{eqnarray}
and
\begin{eqnarray}
  \biggl[
    -\frac{\partial^2}{\partial t^2} + m^2 &+& \frac{\lambda_0}{3!N}\,q^2(t)
    + \frac{g_0}{5!N^2}\,q^4(t) \nonumber \\
    & + &\frac{\lambda_0}{3!}\,G(t,t)
    + \frac{g_0}{5!}\,G^2(t,t) + \frac{2\,g_0}{5!N}\, q^2(t)\,G(t,t)
  \biggr]\, G(t,t') =  \Gamma\,\delta(t-t').
\label{eq:fluct}
\end{eqnarray}
These coupled dynamical equations are exact to leading order in $N$.

The effective dynamical equations (\ref{eq:op}) and (\ref{eq:fluct}) can
be also derived by a variational approach to the path integral
(\ref{eq:gf2}), or Hartree approximation,
where one seeks
for the best quadratic approximation for the action.

\section{The solution}
\label{sec:solution}
To solve the
coupled effective dynamical equations (\ref{eq:op}) and (\ref{eq:fluct})
we rewrite them as
\begin{eqnarray}
    \frac{\partial^2 q(t)}{\partial t^2} &=& F\bigl[ q(t),G(t,t)\bigr]\, q(t)
\label{eq:cinque} \\
    \frac{\partial^2 G(t,t')}{\partial t^2} &=& F\bigl[ q(t),G(t,t)\bigr]\,
       G(t,t') -\Gamma \delta(t-t')
\label{eq:sei}
\end{eqnarray}
where
\begin{eqnarray}
     F\bigl[ q(t),G(t,t)\bigr] = m^2 &+& \frac{\lambda_0}{3!N}\,q^2(t)
    + \frac{g_0}{5!N^2}\,q^4(t) \nonumber \\
    &+& \frac{\lambda_0}{3!}\,G(t,t)
    + \frac{g_0}{5!}\,G^2(t,t) + \frac{2\,g_0}{5!N}\, q^2(t)\,G(t,t)
\label{eq:sette}
\end{eqnarray}
We are eventually interested into solutions for times $t$ such that
$t/\tau\to 0$. Under this assumption the effective dynamical equations
(\ref{eq:cinque}) and (\ref{eq:sei}) can be reduced
to simpler first order
non linear differential equations by using the following representation for
$q(t)$ and $G(t,t')$:
\begin{equation}
\label{eq:uno}
   q(t) =  q(0)\, f_1(t)
\end{equation}
and
\begin{equation}
   G(t,t') =  f_1(t')\, f_2(t)\,  \theta(t'-t) +
              f_1(t)\,  f_2(t')\, \theta(t-t')
\label{eq:due}
\end{equation}
with
\begin{eqnarray}
   f_1(t) &=& {\rm e}^{-\int_0^t d \tau R(\tau)}
\label{eq:tre} \\
   f_2(t) &=& \Gamma\, {\rm e}^{-\int_0^t}\,
                     {\rm e}^{2\int_0^{\tau} d \tau' R(\tau')}
\label{eq:quattro}
\end{eqnarray}
where the function $R(t)$ is solution of the first order non linear
differential equation
\begin{equation}
    \frac{\partial R(t)}{\partial t} =R^2(t) -F\bigl[ q(t),G(t,t)\bigr]
\label{eq:otto}
\end{equation}
By inspection of eqs. (\ref{eq:sette}) and (\ref{eq:otto}) it follows that
$R(t)$ should have the functional form
\begin{equation}
    R(t) = \alpha C(t) +\beta q^2(t)+\gamma
\label{eq:nove}
\end{equation}
where $C(t)=\lim_{t \to t'} G(t,t')$. The parameters $\alpha$, $\beta$
and $\gamma$ are determined by
substituting $R(t)$ from eq. (\ref{eq:nove}) into eq. (\ref{eq:otto}),
and eliminating ${\rm d}C(t)/{\rm d}t$ and ${\rm d}q(t)/{\rm d}t$
with the help of
\begin{eqnarray}
    \frac{\partial q(t)}{\partial t} &=& -R(t) q(t)
\label{eq:dieci} \\
    \frac{\partial C(t)}{\partial t} &=& -2 R(t) C(t) +\Gamma
\label{eq:undici}
\end{eqnarray}
obtained from eqs. (\ref{eq:uno}) - (\ref{eq:quattro}).
We obtain the following solution for $R(t)$
\begin{equation}
\label{eq:dodici}
    R(t) =\frac{\lambda}{6} C(t)+\frac{\lambda}{6N} q^2(t)+a.
\end{equation}
In principle there exists another set of parameters,
but it leads to an asymptotically unsteady state.

The first order non linear differential equations
(\ref{eq:dieci}) - (\ref{eq:dodici}) gives the full description
of the model (\ref{eq:LangN}) - (\ref{eq:GinzLand}) in the
limit $N\to\infty$ for all times $t$ such that $t/\tau\to 0$.

If $q(t)$ is not identically equal to zero it is not straightforward to solve
analytically the set of equations (\ref{eq:dieci}) - (\ref{eq:dodici}).
Nevertheless these can be easily solved numerically for any set of
initial conditions. We note that this is not the case for
eqs. (\ref{eq:op}) and (\ref{eq:fluct}). These, indeed, suffer of
strong numerical instability and one has to resort clever algorithm to
handle them.

On the contrary if $q(t)=0$ for all times, then we can find
a closed analytical solution. We note that
from the structure of the equations on can see that the $O(N)$ symmetry
dictates the equilibrium value of $q(t)$:
$\lim_{t \to \infty}q(t)=0$. Thus if we assume that $q(0)=0$ then the
solution is $q(t)=0$. In this case the equation for $G(t,t')$ can also
be solved, and if we take as initial condition $C(t=0)=0$,
the solution reads
\begin{eqnarray}
 f_1(t) &=& {\rm e}^{-at/2}\,\sqrt{ 1 -(\alpha/\beta) \over
      {\rm e}^{\lambda\Delta t/3} -
       (\alpha/\beta)\, {\rm }e^{-\lambda\Delta t/3}}
\label{eq:tredici} \\
 f_2(t) &=& {3\Gamma\over\lambda\Delta}\,
                 {\rm e}^{at/2}\,\sqrt{ 1 -(\alpha/\beta) \over
      {\rm e}^{\lambda\Delta t/3} -
      (\alpha/\beta)\, {\rm e}^{-\lambda\Delta t/3}} \,
	\sinh(\lambda\Delta t/3)
\label{eq:quattordici}
\end{eqnarray}
where
\begin{equation}
 \alpha = -{3\,a \over \lambda} +
          \sqrt{ \gamma^2 + {3\,\Gamma \over \lambda} } \qquad\qquad
 \beta = -{3\,a \over \lambda} -
          \sqrt{ \gamma^2 + {3\,\Gamma \over \lambda} }
\label{eq:quindici}
\end{equation}
Substitution of eqs. (\ref{eq:tredici}) -- (\ref{eq:quindici}) into eq.
(\ref{eq:due}) leads to the solution for $G(t,t')$. In the limit
$t'\to t$ we recover the result of Ref. \cite{Scalapino} for $C(t)$.

\section{Beyond the Hartree approximation}
\label{sec:bHartree}
The next leading terms of order $1/N$ can be included systematically by
evaluating diagrams not included in eq. (\ref{eq:g2n}). In the case
$q(t)= 0$ the work is simplified since one does not have to
consider separately transverse and parallel components of
the correlation function
$\langle\phi_i(t)\,\phi_j(t')\rangle_{\rm c}$.

The diagrams contributing to the first corrections to $\Gamma_2$ are shown
in Fig.~\ref{fig:fig1} (a) and (b) and Fig.~\ref{fig:fig2} and yields
\begin{eqnarray}
   \Gamma_2(G) = &&\frac{2}{4!}\lambda_0\int{\rm d}t\, G^2(t,t)
                     + \frac{6}{6!}g_0\,\int{\rm d}t\, G^3(t,t)
                     \nonumber \\
           &-& \frac{\lambda_0^2}{12^2}\,\int{\rm d}t\,\int{\rm d}t'\,
                G^4(t,t')
             - \frac{36}{(6!)^2}g_0^2\,\int{\rm d}t\,\int{\rm d}t'\,
                     G^4(t,t')\, G(t,t)\, G(t',t')
                     \nonumber \\
             &-& \frac{\lambda_0\,g_0^2}{720}\,\int{\rm d}t\,\int{\rm d}t'\,
                     G^4(t,t')\, G(t,t)
\label{eq:g2cor}
\end{eqnarray}

To improve systematically this result, and the Hartree approximation, one has
to consider an infinite series of diagrams.
A complete summation of the series, see Figs.~\ref{fig:fig3} and
\ref{fig:fig4}, can be performed, however,  only in
the case of a system at equilibrium where time translational invariance
holds. The final result for $\Gamma_2$ is valid to all orders in
$\lambda$ and to first order in $1/N$ and reads \cite{Bray,Townsend}:
\begin{eqnarray}
   \Gamma_2(G) / \tau = &&\frac{N+2}{4!} \lambda_0
                         \int \frac{{\rm d} \omega_1}{2 \pi}
                         \int \frac{{\rm d} \omega_2}{2 \pi}
                         \widetilde G(\omega_1) \widetilde G(\omega_2)
                         \nonumber \\
                        &+&\frac{N+6}{6!} g_0
                          \int \frac{{\rm d} \omega_1}{2 \pi}
                          \int \frac{{\rm d} \omega_2}{2 \pi}
                          \int \frac{{\rm d} \omega_3}{2 \pi}
                          \widetilde G(\omega_1) \widetilde G(\omega_2)
                          \widetilde G(\omega_3)
                          \nonumber \\
                        &+&\frac{1}{2}
                          \int \frac{{\rm d} \omega}{2 \pi}
                          \ln\left[1 +\left(\frac{\lambda_0}{3!}+
                                            \frac{g_0}{60}\,G(0,0)\right)
                          \widetilde \Pi(\omega)\right]
                          \nonumber \\
                        &-&\frac{1}{2}
                          \int \frac{{\rm d} \omega}{2 \pi}
                  \left(\frac{\lambda_0}{3!}+ \frac{g_0}{60}\,G(0,0)\right)
                          \widetilde \Pi(\omega))
\label{eq:g2cor2}
\end{eqnarray}
where $\widetilde \Pi(\omega)$ is the so called vacuum polarization propagator:
\begin{equation}
\label{eq:polar}
\widetilde \Pi(\omega)= \int \frac {{\rm d} \omega}{2 \pi}\,
                        \widetilde G(\eta)\, \widetilde G(\eta+\omega).
\end{equation}
Upon differentiating with respect to $\widetilde G(\omega)$ the functional
$\Gamma[G]$ with the $1/N$ corrections included one obtains \cite{Dominici}:
\begin{eqnarray}
  \widetilde{G}^{-1}(\omega)= \omega^2 &+& m^2 + \frac{\lambda_0}{3!}\,G(0,0)
                               + \frac{g_0}{5!}\,G^2(0,0)
                               \nonumber \\
                   & + &\frac{1}{N}
                         \int \frac{{\rm d} \omega_1}{2 \pi}
      \frac{1}{\left[1 +\left(\frac{\lambda_0}{3!}+
\frac{g_0}{60}\,G(0,0)\right)
       \widetilde \Pi(\omega_1)\right]}
       \nonumber \\
  &&\phantom{xxxxxxxxxxxxx}
   \times \left[\left(\frac{\lambda_0}{3}+ \frac{g_0}{30}\,G(0,0)\right)
   \widetilde G(\omega-\omega_1)+\widetilde \Pi(\omega_1)\right]
\label{eq:equil}
\end{eqnarray}
Equation (\ref{eq:equil})
represents the spectrum of the equilibrium fluctuations correct to
order $1/N$. The study of these corrections will be the subject of a
future publication.

\section{Variational Approach}
\label{sec:variational}
A well-known approximation often employed in statistical mechanics
is obtained by applying the Peierls-Feynman-Bogolubov inequality:
\begin{equation}
\label{eq:appe1}
-\ln Z \leq -\ln Z_0 +\langle{\cal L}-{\cal L}_0\rangle_0=W^{(1)}
\end{equation}
where ${\cal L}$ is action density given by
\begin{equation}
   I(\phi) = \int_{0}^{\tau}\, {\rm d} t\, {\cal L}
\label{eq:acden}
\end{equation}
and ${\cal L}_0$ is an arbitrary action density. The average
$\langle\cdots\rangle_0$ in (\ref{eq:appe1}) is done with
respect to the probability distribution corresponding
to ${\cal L}_0$ and $Z_0$ is the partition function associated with
${\cal L}_0$.
The Hartree method consists of choosing the arbitrary action ${\cal L}_0$
to be the most general quadratic form:
\begin{equation}
\label{eq:appe2}
   \int_{0}^{\tau} {\rm d}t\, {\cal L}_0 = \frac{1}{2 \Gamma}
                 \int_{0}^{\tau} {\rm d}t\, \int_{0}^{\tau} {\rm d}t'\,
                 \sum_{i,j} \phi_i(t)\, u_{ij}(t,t')\, \phi_j(t')
               - \int_{0}^{\tau} {\rm d}t\, \sum_i v_i(t)\, \phi_i(t)
\end{equation}
where $u_{ij}$ is a positive definite kernel.
Recalling that:
\begin{eqnarray}
       W^{(0)}&=& -\ln Z_0 \nonumber \\
              &=& -\frac{1}{2}\, \ln||u^{-1}(t,t')||
                  -\frac{\Gamma}{2}\int_{0}^{\tau} {\rm d} t
                                   \int_{0}^{\tau} {\rm d} t'
                                     v(t)\,u^{-1}(t,t')\,v(t')
\label{eq:w0}
\end{eqnarray}
Taking the derivatives of $W^{(0)}$ with respect to $v(t)$ we find:
\begin{eqnarray}
\frac{\delta W^{(0)}}{\delta v(t)}&=& -q(t)
                                   =-\Gamma \int_{0}^{\tau} {\rm d}t'
                                          u^{-1}(t,t')\,v(t')
\label{eq:d1} \\
\frac{\delta^2 W^{(0})}{\delta v(t)\, \delta v(t')}&=& -G(t,t')
                                                    = -\Gamma\,  u^{-1}(t,t')
\label{eq:d2}
\end{eqnarray}
which can be inverted to give:
\begin{eqnarray}
       \int_{0}^{\tau} {\rm d}t'\, u(t,t')\, q(t') &=&\Gamma\, v(t)
\label{eq:appe3} \\
        \int_{0}^{\tau} {\rm d}t'\, u(t,t')\,G(t',t'') &=&\Gamma\,
\delta(t-t'')
\label{eq:appe4}
\end{eqnarray}
The arbitrary functions $u_{ij}$ and $v_i$ are determined by looking for
the minimum of $W^{(1)}$ to have the best estimate of $\ln Z$.

For the case under study by applying the above method
to (\ref{eq:GinzLand}) one finds
\begin{eqnarray}
  u(t,t')=\bigl[
    -\frac{\partial^2}{\partial t^2} + m^2 &+& \frac{\lambda_0}{3!N}\,q^2(t)
    + \frac{g_0}{5!N^2}\,q^4(t)
    \nonumber \\
    &+& \frac{\lambda_0}{3!}\,G(t,t)
    + \frac{g_0}{5!}\,G^2(t,t) + \frac{2\,g_0}{5!N}\, q^2(t)\,G(t,t)
  \bigr]\, \delta(t-t').
\label{eq:appe5}
\end{eqnarray}
and
\begin{equation}
\label{eq:appe6}
     v(t)=0
\end{equation}
Thus inserting (\ref{eq:appe5}) and (\ref{eq:appe6}) in eqs.
(\ref{eq:appe3}) and (\ref{eq:appe4}) we find
the same result as eqs. (\ref{eq:op}) and (\ref{eq:fluct}).

Before concluding the discussion of the Hartree method we shall
analyze what happens in the case $N=1$.
Its variational nature in fact justifies its application
even when the problem under scrutiny does not
contain a natural small parameter around which to perform some
sort of expansion.

Let us consider the case of a Langevin equation
for a scalar field with cubic nonlinearity
The equations of motion for the average value of the field and
for the fluctuations are \cite{Umberto}:
\begin{eqnarray}
  \bigl[
    -\frac{\partial^2}{\partial t^2} + m^2 &+& \frac{\lambda_0}{3!}\,q^2(t)
    + \frac{g_0}{5!}\,q^4(t)
    \nonumber \\
   &+&\frac{\lambda_0}{2}\,G(t,t)
    + \frac{g_0}{8}\,G^2(t,t) + \frac{g_0}{12}\, q^2(t)\,G(t,t)
  \bigr]\, q(t) = 0
\label{eq:op1} \\
  \bigl[
    -\frac{\partial^2}{\partial t^2} + m^2 &+& \frac{\lambda_0}{2}\,q^2(t)
    + \frac{g_0}{4!}\,q^4(t)
    \nonumber \\
    &+& \frac{\lambda_0}{2}\,G(t,t)
    + \frac{g_0}{8}\,G^2(t,t) + \frac{g_0}{8}\, q^2(t)\,G(t,t)
  \bigr]\, G(t,t') =  \Gamma\,\delta(t-t').
\label{eq:fluct1}
\end{eqnarray}
With the substitution:
\begin{equation}
q(t)=e^{-\int_0^t d \tau R_q(\tau)}
\end{equation}
and $G(t,t')$ given by eq. (\ref{eq:due})
we obtain two equations for $R_q(t)$
and $R(t)$ analogous to eq. (\ref{eq:otto}). Using the trial solution
\begin{equation}
R_q(t)=\alpha_q\, C(t)  + \beta_q\, q^2(t) +\gamma_q
\end{equation}
and $R(t)$ given by  eq.(\ref{eq:nove})  we find,
up to terms quadratic in the coupling constant $\lambda$,
 $\alpha_q=\alpha=\lambda/2$,
$\beta_q=\lambda/6$, $\beta=\lambda/2$ and $\gamma_q=\gamma=a$.
The value of the coefficients coincides with the value obtained
in the so called Langer-Bar on-Miller approximation \cite{Langer}
to the Langevin
equation, a result which was also rediscovered few years ago
\cite{Indiani}, on the basis of a somehow ad hoc
variational principle.
We believe that the present derivation, being based on a path integral
formulation of the stochastic equations makes the underlying physical
assumptions more clear.

\section{Conclusions}
\label{sec:conclusion}
Most problems arising in the study of physical problems are most naturally
represented in terms of systems of nonlinear stochastic differential
equations. Many approximation schemes have been developed to treat the
nonlinear aspect of equations. Usually they are based on some reasonable
assumptions. The advantage of these approaches is that they may lead to
relative simple equations. The drawback is that is is not simple to
improve the quality of the approximation. In this paper we presented
an alternative approach to the study of nonlinear Langevin equation which
allows for systematic development of approximation scheme. The basic idea
is to reduce the nonlinear Langevin equation to an equivalent equilibrium
problem to which the methods of conventional field theory can be applied.
A particular well suited perturbative scheme is that developed in quantum
field theory by Cornwall, Jackiw and Tomboulis \cite{Cornwall}. The major
advantage is that it leads to a variational principle for the physical
quantities of interest.

The method is applied to an $N$ component Ginzburg-Landau equation. In the
limit of $N\to\infty$ we are able to derive closed forms for the order
parameter $q(t)$ and for the two-time connected correlation function $G(t,t')$.
We also discuss the first $1/N$ corrections. The study of these is,
however, more involved and has not been included in this paper. This is part
of future work.
It will be also of interest to extend the present approach to higher dimensions
and to explore numerically the predictions of the present approach
to finite values of $N$.

\acknowledgments
We thank D. Dominici and A. Maritan for useful discussions

\begin{figure}
\caption{Leading order 2PI diagrams}
\label{fig:fig1}
\end{figure}

\begin{figure}
\caption{2PI diagrams contribution to the first $1/N$ corrections}
\label{fig:fig2}
\end{figure}

\begin{figure}
\caption{Three vertices 2PI diagrams contributing to the first $1/N$
          corrections
         }
\label{fig:fig3}
\end{figure}

\begin{figure}
\caption{Four vertices 2PI diagrams contributing to the first $1/N$
         corrections
         }
\label{fig:fig4}
\end{figure}

\end{document}